\documentstyle[12pt]{article}
\begin{document} 
\vspace*{-1in} 
\renewcommand{\thefootnote}{\fnsymbol{footnote}} 
\begin{flushright} 
TIFR/TH/99-01\\
IMSc/99/01/03\\
January 1999\\ 
hep-ph/9901276 
\end{flushright} 
\vskip 65pt 
\begin{center} 
{\Large \bf $J/\psi+\gamma$ production at the LHC}\\
\vspace{8mm} 
{\bf 
Prakash Mathews\footnote{prakash@theory.tifr.res.in}, 
K.~Sridhar\footnote{sridhar@theory.tifr.res.in}
}\\ 
\vspace{10pt} 
{\sf
Department of Theoretical Physics, Tata Institute of 
Fundamental Research,\\  
Homi Bhabha Road, Bombay 400 005, India. } 

\vspace{25pt} 
 
{\bf Rahul Basu\footnote{rahul@imsc.ernet.in}\\   }
\vspace{10pt} 
{\sf Institute of Mathematical Sciences, CPT Campus, Chennai 600 113, India.}

\vspace{80pt} 
{\bf ABSTRACT} 
\end{center} 
\vskip12pt 
The associated production of $J/\psi + \gamma$ at the LHC is studied
within the NRQCD framework. The signal we focus on is the production
of a $J/\psi$ and an isolated photon produced back-to-back, with
their transverse momenta balanced. It is shown that even for very large
values of transverse momentum ($p_T \sim 50$~GeV) the dominant 
contribution to this process is $not$ fragmentation. This is because
of the fact that fragmentation-type contributions to the cross-section
come from only a $q \bar q$ initial state, which is suppressed at the
LHC. We identify $gg$-initiated diagrams higher-order in $\alpha_s$ 
which do have fragmentation-type vertices.
We find, however, that the contribution of these diagrams is
negligibly small.
\setcounter{footnote}{0} 
\renewcommand{\thefootnote}{\arabic{footnote}} 
 
\vfill 
\clearpage 
\setcounter{page}{1} 
\pagestyle{plain}
\noindent NRQCD is an effective field theory derived from the full QCD
Lagrangian by neglecting all states of momenta larger than
a cutoff of the order of the heavy quark mass, $m$ \cite{bbl}, and 
accounting for this exclusion by introducing new interactions in the 
effective Lagrangian, which are local since the excluded states
are relativistic. It is then possible to expand the quarkonium state
in terms of its Fock-components as a perturbation series in $v$ (where
$v$ is the relative velocity between the heavy quarks), and
in this expansion, the $Q \bar Q$ states appear in either
colour-singlet or colour-octet configurations.
The colour-octet $Q \bar Q$ state is connected to the physical
state by the emission of one or more soft gluons by transitions
which are dominantly non spin-flip and spin-flip transitions.
Selection rules for these radiative transitions then allow us
to keep track of the quantum numbers of the octet states, so that
the production of a $Q \bar Q$ pair in a octet state
can be calculated and its transition to a physical singlet state
can be specified by a non-perturbative matrix element. The cross-section
for the production of a meson $H$ then takes on the following factorised form:
\begin{equation}
   \sigma(H)\;=\;\sum_{n=\{\alpha,S,L,J\}} {F_n\over m^{d_n-4}}
       \langle{\cal O}^H_\alpha({}^{2S+1}L_J)\rangle
\label{e1}
\end{equation}
where $F_n$'s are the short-distance coefficients and ${\cal O}_n$ are local
4-fermion operators, of naive dimension $d_n$, describing the long-distance
physics. The short-distance coefficients are associated with the production
of a $Q \bar Q$ pair with the colour and angular momentum quantum numbers
indexed by $n$. These involve momenta of the order of $m$ or larger and
can be calculated in a perturbation expansion in the coupling $\alpha_s(m)$.
The non-perturbative long-distance factor $\langle O^H_n\rangle$ is 
proportional to the probability for a pointlike $Q \bar Q$ pair in the 
state $n$ to form a bound state $H$. These matrix elements 
are universal in the sense that having extracted them in a particular 
process, they can be used to make predictions for other processes
involving quarkonia.

In fact, the importance of the colour-octet components was first noted
\cite{bbl2} in the case of $P$-wave charmonium decays, and even in
the production case the importance of these components was first
seen \cite{jpsi} in the production of $P$-state charmonia at the
Tevatron. The surprise was that even for the production of $S$-states
such as the $J/\psi$ or $\psi'$, where the colour singlet components give
the leading contribution in $v$, the inclusion of sub-leading octet states
was seen to be necessary for phenomenological reasons \cite{brfl}. 
While the inclusion of the colour-octet components seem to be
necessitated by the Tevatron charmonium data, the normalisation of
these data cannot be predicted because the long-distance matrix 
elements are not calculable. The data allow a linear combination of 
octet matrix-elements to be fixed \cite{cgmp,cho}, and much effort
has been made recently to understand the implications of these colour-octet
channels for $J/\psi$ production in other processes: for example, $J/\psi$
production at the LEP \cite{lep5,lep6}, the prediction for the polarisation
of the $J/\psi$ \cite{trans}, production of $J/\psi$ at fixed-target 
$pp$ and $\pi p$ experiments \cite{hadro}, inelastic photoproduction 
at HERA \cite{cakr,brw,sms}, production of quarkonium states at the 
LHC \cite{lhc} and the predictions for the large-$p_T$ production of 
other charmonium resonances at the Tevatron \cite{ours}. Recently,
next-to-leading order calculations for quarkonium production at
low-$p_T$ have also been completed \cite{mangano} and make it
possible to make accurate predictions for these processes.

%=======================================================================%
\begin{figure}[ht]
\vskip 0.4in\relax\noindent
          \relax{\includegraphics{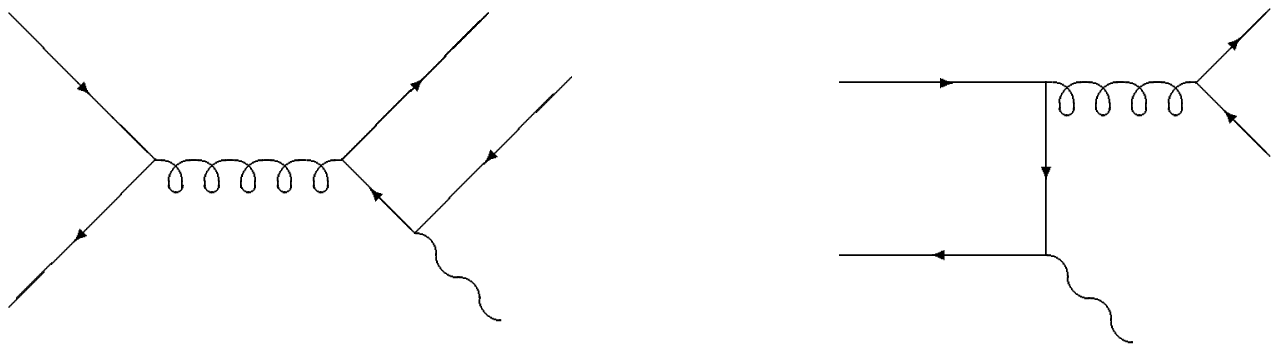}}  
\vskip 1in\relax\noindent
\caption{Quark tree diagrams.}
\end{figure}
%=======================================================================%

In this paper we consider the large $p_T$ associated production of 
an isolated photon
and $J/\psi$ produced back-to-back at LHC energies. This process was first
studied \cite{psigam} in the context of the colour-singlet model and more
recently, the
contribution of the colour-octet channels at the Tevatron has also been
studied \cite{kim,us}. The tree-level cross-section for the $J/\psi + \gamma$
process can be obtained from the corresponding cross-sections for the
photoproduction of $J/\psi$ which have been calculated in Ref.~\cite{cakr}.
It turns out that, because of the photon in the final state, the 
colour-singlet contribution is more important for this process than
for inclusive $J/\psi$ production at the Tevatron. In fact, this process
can be a sensitive probe of quarkonium production and it is our aim, in
the present work, to understand how it may be used to unravel aspects
of quarkonium production dynamics at the LHC. 

%=======================================================================%
\begin{figure}[ht]
\vskip 0.4in\relax\noindent
          \relax{\includegraphics{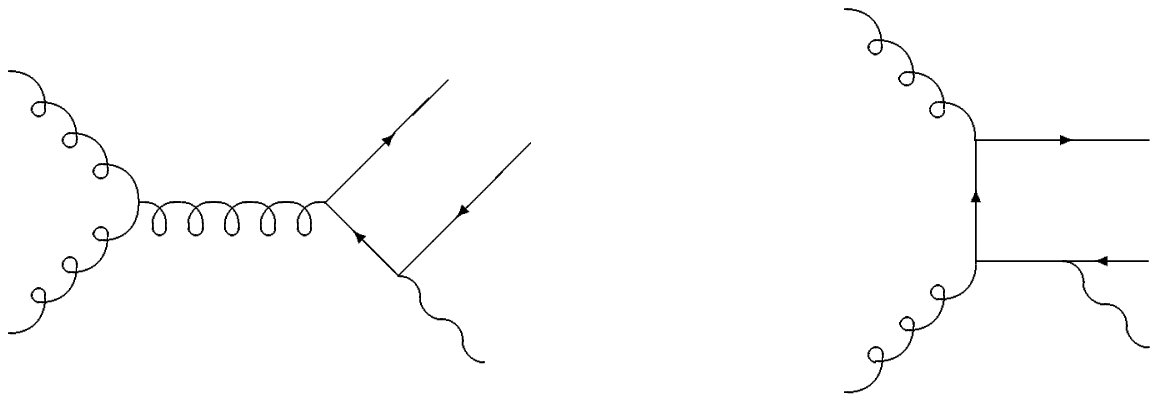}}  
\vskip 1in\relax\noindent
\caption{Gluon tree diagrams.}
\end{figure}
%=======================================================================%

The contributing subprocesses to $J/\psi + \gamma$ production are
\begin{eqnarray}
q ~\bar q ~ \rightarrow ~ ^{2S+1}L_J ~\gamma , \nonumber \\
g ~     g ~ \rightarrow ~ ^{2S+1}L_J ~\gamma~, 
\label{e2}
\end{eqnarray}
The Fock-components that contribute to $J/\psi$ production are
the colour-singlet ${}^3S_1^{[1]}$ state and the colour-octet states
${}^3S_1^{[8]}$, ${}^1S_0^{[8]}$ and ${}^3P_{0,1,2}^{[8]}$. 
The colour-singlet ${}^3S_1$ state contributes at ${\cal O}(1)$ but the
colour-octet channels all contribute higher orders in $v$. This is 
because the ${}^3S_1^{[8]}$ connects to the $J/\psi$ by emitting
two soft gluons (both non spin-flip transitions and
resulting in an effective ${\cal O}(v^4)$ suppression)
while the ${}^1S_0^{[8]}$ connects to the physical state via a spin-flip 
transition (the correct power-counting for which yields an 
an effective ${\cal O}(v^3)$ suppression \footnote{See the erratum of 
Ref.~\cite{bbl}}).
On the other hand, the ${}^3P_{0,1,2}^{[8]}$ states connect to
the $J/\psi$ by a single non spin-flip transition but since they are $L=1$
states their production cross-section is already suppressed by
${\cal O}(v^2)$, making them effectively of ${\cal O}(v^4)$.

%============================================================================%
\begin{figure}[ht]
\vskip 0.5in\relax\noindent
          \relax{\includegraphics{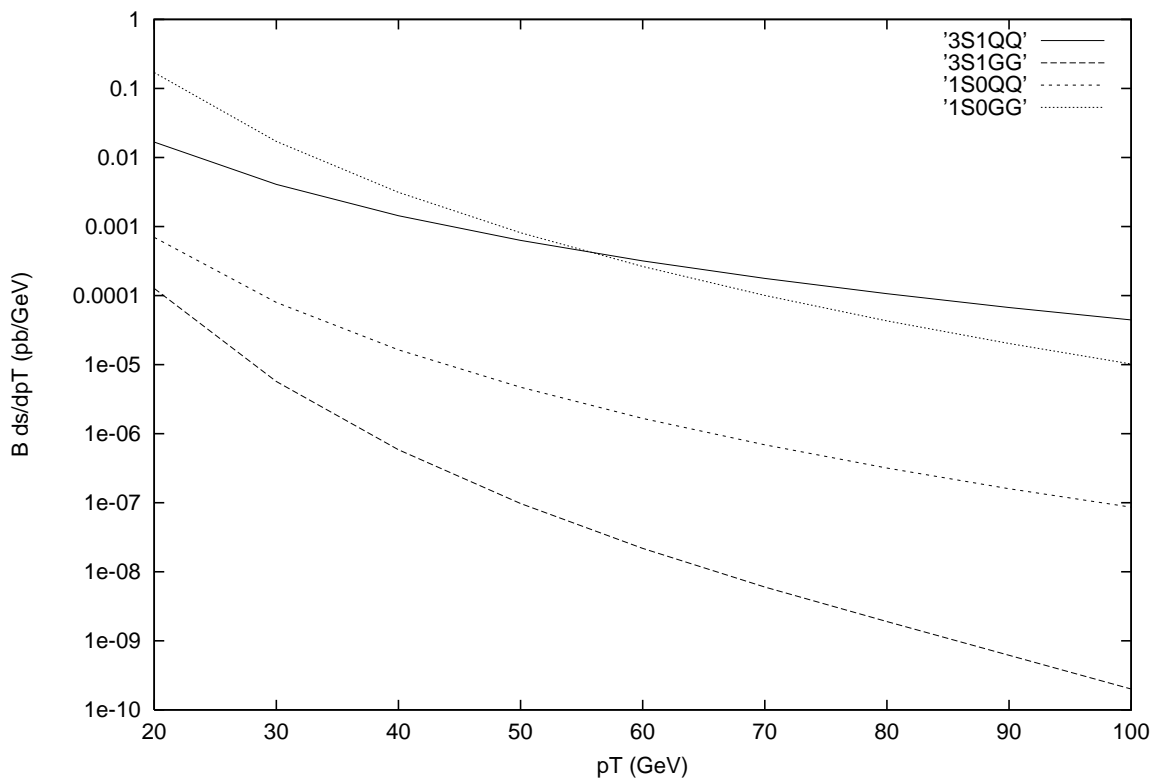}}  
\vskip 3.8in\relax\noindent
\caption{$B d\sigma/dp_T$ for $J/\psi+\gamma$ 
production at the LHC. } 
\end{figure}
%============================================================================%

In the $\alpha_s$ perturbation series,
we will first study all contributions to ${\cal O}(\alpha \alpha_s^2)$
i.e. the tree-level diagrams, and then we will study the effect of
a class of higher-order (in $\alpha_s$) diagrams which are likely
to be important for the process under consideration.
The net contribution of the various subprocess depends on various factors:
\begin{itemize} 
\item
the initial parton flux
\item
the order in $v$ of the final contributing non-perturbative NRQCD
matrix element, and  
\item
the $p_T$ behaviour of the subprocess.
\end{itemize} 

%=======================================================================%
\begin{figure}[ht]
\vskip0.7in\relax\noindent
          \relax{\includegraphics{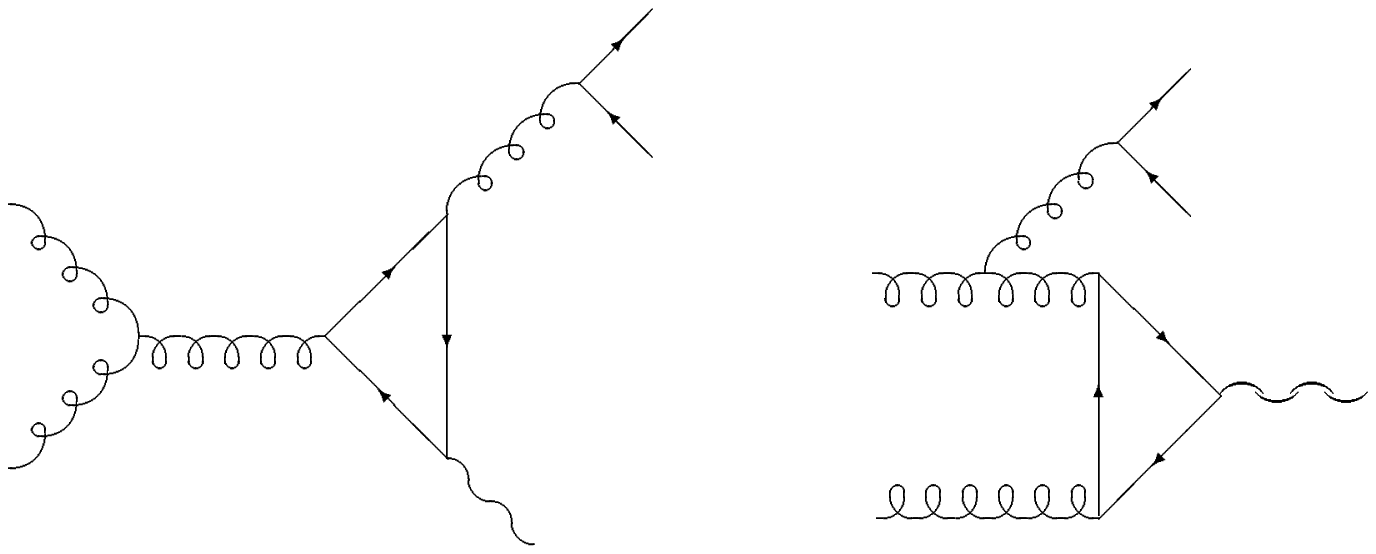}}  
\vskip 1in\relax\noindent
\caption{Fragmentation type triangle diagrams.}
\end{figure}
%=======================================================================%

In Figs.~1 and 2, we have shown some examples of tree-level
contributions to this process. From our experience with large-$p_T$ 
$J/\psi$ production at the Tevatron, one gleans the fact that is
expedient to classify these diagrams in terms of the number of heavy-quark
propagators in them. To this end we introduce some notation: the diagrams
where only one gauge boson attaches itself to the $Q\bar Q$ pair is
called a one-vertex or 1V diagram. Such diagrams have no heavy-quark
propagators and are fragmentation-like diagrams. Similarly there are
two-vertex (2V) and three-vertex (3V) diagrams, which have one and two
heavy quark propagators, respectively. This classification is useful
in understanding the systematics of $J/\psi$ production at large-$p_T$
at the Tevatron. In the case of inclusive production, it turns out
that the 1V diagrams (which are fragmentation-like) are those that
dominate at large-$p_T$. It is straightforward to convince oneself
(by using the spin and colour projectors used in quarkonium calculations)
that the only state that can be produced in a 1V diagram is the 
${}^3S_1^{[8]}$ state. Consequently, $J/\psi$ production at large
$p_T$ is completely dominated by ${}^3S_1^{[8]}$ production \cite{cho}. 
On the other hand, at the lower end of the $p_T$ spectrum at the
Tevatron, there is a significant contribution from the 
${}^1S_0^{[8]}$ and ${}^3P_{0,1,2}^{[8]}$ states. These come
from 2V diagrams with the contribution from the 3V diagrams being
less important. In fact, it is this $p_T$ dependence that allows
for a separate determination of the $\langle {}^3S_1^{[8]}\rangle $ 
matrix element. On the other hand, it is not possible to determine the 
$\langle {}^1S_0^{[8]}\rangle $ and $\langle {}^3P_{0,1,2}^{[8]}\rangle $ 
matrix elements separately because they have a similar $p_T$ dependence
owing to the fact that the production of these states proceed dominantly
via 2V diagrams. 

%=======================================================================%
\begin{figure}[ht]
\vskip0.3in\relax\noindent
          \relax{\includegraphics{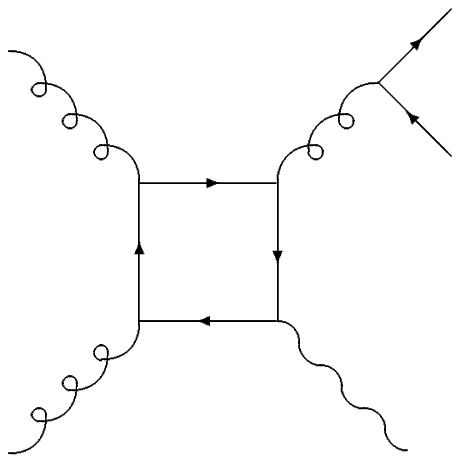}}  
\vskip 1.2in\relax\noindent
\caption{Box digram.}
\end{figure}
%=======================================================================%

At the LHC, the situation is somewhat different. At
the large energies and large $p_T$ values that will be available at
the LHC, the production of $J/\psi$ is expected to be overwhelmed
by the intermediate production of a ${}^3S_1^{[8]}$ state. The contribution
of the ${}^1S_0^{[8]}$ and ${}^3P_{0,1,2}^{[8]}$ states is very small
\cite{lhc}. The complete dominance of the fragmentation-type contributions
for $J/\psi$ production at the LHC is also because of the fact that the
production mechanisms in LHC which dominate are $gg$-initiated. This
is an important point to keep in mind when we study $J/\psi + \gamma$
production, because if we consider the tree diagrams for this process
we find that there is no 1V diagram contributing through a $gg$-initiated
channel. The only 1V diagram that contributes is in the $q \bar q$-initiated
channel. Consequently, we expect that fragmentation-type sub-processes are
not as important to this process as they are for the case of $J/\psi$
production. 

To check out these expectations, we have studied the production of 
$J/\psi+\gamma$ at the LHC ($\sqrt{s}=14$~TeV). In Fig.~3, we present
the results for the cross-section $Bd\sigma/dp_T$ as a function of
$p_T$. We have assumed $-2.5<y<2.5$ in our computations. For the
input parton distributions, we use the MRSD-$^\prime$ \cite{mrs}
and evolved them to the scale $Q=M_T$.  
For the numerical values of the relevant non-perturbative matrix elements 
we use the numbers tabulated below which have been obtained \cite{cho} 
by fitting to the CDF data. We would like to point out here that the
inclusion of soft-gluon radiation effects \cite{ccsl, kniehl, sms}
lead to lower fitted values for the non-perturbative parameters,
but these effects are somewhat model-dependent. For the purposes
of the present analysis, we prefer to use the matrix elements derived
by Ref.~\cite{cho}.
\begin{eqnarray}
\left<{\cal O}^{J/\psi}_1 (^3 S_1)\right> &=& 1.2 ~{\rm GeV^3}~,\nonumber \\
\left<{\cal O}^{J/\psi}_8 (^3 S_1)\right> &=& (6.6 \pm 2.1) 
\times 10^{-3} ~{\rm GeV^3~,}\\
\frac{\left<{\cal O}^{J/\psi}_8 (^3 P_0)\right>}{M_c^2}+ 
\frac{\left<{\cal O}^{J/\psi}_8 (^1 S_0)\right>}{3} &=& 
(2.2 \pm 0.5) \times 10^{-2} ~{\rm GeV^3}~. \nonumber
\label{e3}
\end{eqnarray}
Since only the sum of $\langle {}^1 S_0^{[8]}\rangle $ and 
$\langle {}^3 P_0^{[8]}\rangle $ matrix elements
can be extracted from the $J/\psi$ CDF data, we make prediction by
considering the maximal case, $i.e.$ saturating the sum by either the 
$^1 S_0^{[8]}$ or $^3 P_0^{[8]}$ matrix elements.  Using heavy quark spin 
symmetry the other matrix elements are related $\left<{\cal O}^{J/\psi}_8 
(^3 P_J)\right>= (2 J+1) \left<{\cal O}^{J/\psi}_8 (^3 P_0)\right>$. 
First let us consider the case in which the 
$\langle {}^1 S_0^{[8]}\rangle$ saturates the sum in Eq.~3.  
The contributions 
would hence come from the $^1 S_0^{[8]}$ and the $^3 S_1^{[8]}$ terms.  
We find that even upto a value of 50~GeV in $p_T$, the dominant
contribution is that which comes from the 2V diagrams involving the
$^1 S_0^{[8]}$ state. This is because of the fact that this subprocess
is $gg$ initiated. It is only above a $p_T$ of 50 GeV that the 
$q \bar q$-initiated
1V diagram starts dominating. The upshot of this computation is that 
at the LHC energies there is an interesting interplay between the initial parton
flux and the fragmentation-type effects, and because of the fact that the
$q \bar q$ flux is small at these energies, the effects of fragmentation
do not show up till the $p_T$ values become very large. We have also checked
that the results are more or less unchanged even if we saturate the sum in
Eq.~3 with the $\langle {}^3P_{0}^{[8]}\rangle $ matrix element.

It is also
important to check whether this is only because we have restricted to
tree-level diagrams. In principle, it is possible that 1V diagrams 
which are higher order in $\alpha_s$ and which come from a $gg$ initial 
state could modify this result. In spite of being higher-order in
$\alpha_s$, being 1V diagrams these are possibly enhanced by powers of 
$p_T/m$. In the following we identify the possible 
higher-order diagrams that can contribute and try to estimate the magnitude
of these corrections.

%============================================================================%
\begin{figure}[ht]
\vskip 0.5in\relax\noindent
          \relax{\includegraphics{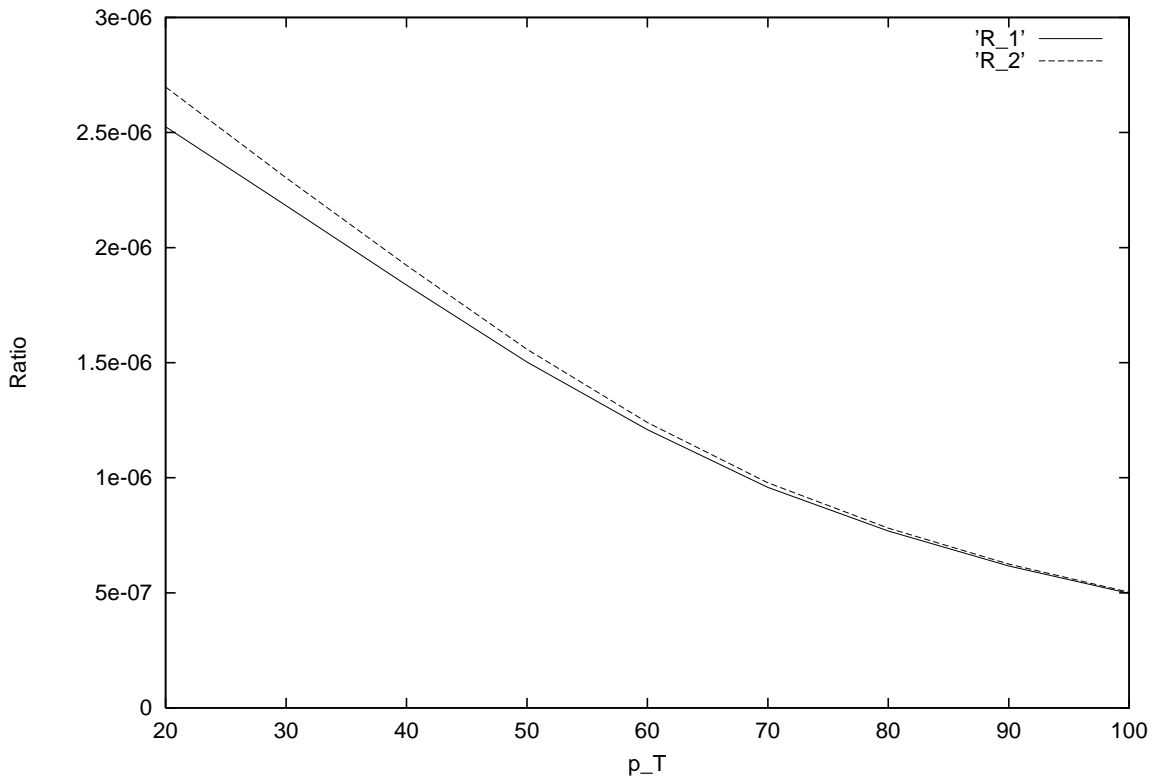}}  
\vskip 3.8in\relax\noindent
\caption{The ratio $R$ of the box diagram contribution
to the tree level cross-section, for the two cases described in the
text.}
\end{figure}
%============================================================================%

Let us consider the possible higher-order diagrams coming from a $gg$
initial state that could contribute to the signal. Since the signal
that we demand is a $p_T$-balanced $J/\psi$ and an isolated photon final 
state,
the only 1V diagrams that we can have are triangle and box diagrams
with quark loops. These are shown in Fig.~4.
It can be showed that the contribution of the 1V triangle 
diagram is zero.  The triangle is attached to one photon and two gluons, 
and vanishes due to Furry's theorem.  

The box diagram (shown in Fig.~5) corresponds to the process $g~g 
\rightarrow g~\gamma$
through a quark loop and the outgoing gluon fragmenting into a $Q\bar Q$
pair which finally forms a $J/\psi$ through an intermediate
$^3 S_1^{[8]}$ state.  
We choose the number of quark flavour in the loop as four.  
This diagram has a non-vanishing interference
with the tree level $g~g \rightarrow Q\bar Q(^3 S_1^{[8]})
~\gamma$.  We begin by evaluating the interference term.  The
box diagram has three terms (corresponding to the three different
ways that the photon can be attached to the internal quark lines)
and three terms with the loop momenta reversed. It turns out that the
contribution of the latter three diagrams are the same as the former. 
The individual diagrams are superficially
divergent and we use dimensional regularisation to regulate it.  Feynman 
parametrisation has to be done symmetrically and leads to a lot of 
simplification.  This gives terms proportional to $(L^2)^n$ where $L$ is 
the loop momenta and n=0,1,2.  Terms proportional to $n=0,1$ are finite
but the $n=2$ term is divergent.  It can be shown that on combining all 
the three diagrams, the $1/\epsilon$ pole cancels to give a finite
part and a logarithmic term. We have extensively used FORM and MATHEMATICA
for the calculation of the interference contribution and use the output
of these packages directly for our numerical computations.  The results
of our computation is shown in Fig.~6, where we have plotted the
ratio $R$ of the magnitude of the interference term to the tree
level cross-section. As before, the results are obtained using
MRSD-' densities and a rapidity cut $-2.5<y<2.5$. The two curves
$R_1$ and $R_2$ shown in Fig.~6 correspond to the two values of
the tree-level cross-sections obtained from saturating the sum
in Eq.~3 with either of the two non-perturbative matrix
elements. We find that the contribution of the interference is tiny, and
the factors associated with the box diagram suppress any 
possible enhancement expected from the gluon flux and fragmentation 
contribution. Given that the interference term is so small, we
expect that the box amplitude square contribution to be
even further suppressed. 

In summary, we have studied the associated production of $J/\psi+\gamma$,
produced back-to-back, 
at the LHC. We find that this process gives us crucial insights into
the dynamics of quarkonium production. In particular, we find that
production via fragmentation-like diagrams does not dominate the 
cross-section upto values of 50~GeV. This is because of the fact 
that there is no such contribution in the $gg$-initiated channel,
but only in the $q \bar q$-initiated channel, where the corresponding
parton flux is small. Beyond the tree level, we find that there is a
box diagram with a $gg$ initial state that contributes to this process,
but we find that the contribution of this diagram to be negligibly small.
 
\vspace{.5cm}
\noindent
{\large \bf Acknowledgments:}
The authors wish to thank the organisers of the Fifth Workshop on
High Energy Physics Phenomenology (WHEPP-5) held in IUCAA, Pune, India 
in January 1998, where this work was initiated. 
\vfill
\eject


\begin{thebibliography}{99}

\bibitem{bbl} G.T.~Bodwin, E.~Braaten and G.P.~Lepage, {\it Phys. Rev.} 
{\bf D 51}, 1125 (1995); erratum $ibid.$ {\bf D 55} 5853 (1997).

\bibitem{bbl2} G.T.~Bodwin, E.~Braaten and G.P.~Lepage, {\it Phys. Rev.} 
{\bf D 46} R1914 (1992). 

\bibitem{cdf}F. Abe {\it et. al}, {\it Phys. Rev. Lett.} {\bf 71}, 
2537 (1993). 

\bibitem{jpsi} E.~Braaten, M.A.~Doncheski, S.~Fleming and M.~Mangano,
{\it Phys. Lett.} {\bf B 333} 548 (1994); D.P.~Roy and K.~Sridhar, 
{\it Phys. Lett.} {\bf B 339} 141 (1994); M.~Cacciari and M.~Greco, 
{\it Phys. Rev. Lett.} {\bf 73} 1586 (1994). 

\bibitem{brfl} E.~Braaten and S.~Fleming, {\it Phys. Rev. Lett.}
{\bf 74} 3327 (1995).

\bibitem{cgmp} M.~Cacciari, M.~Greco, M.~Mangano and A.~Petrelli,
{\it Phys. Lett.} {\bf B 356} 553 (1995).  

\bibitem{cho}
   P. Cho and A.K.~Leibovich, {\it Phys. Rev.}, {\bf D 53} 150 (1996);
   {\it Phys. Rev.}, {\bf D 53} 6203 (1996).

\bibitem{lep5} K.~Cheung, W.-Y.~Keung and T.C.~Yuan, 
{\it Phys. Rev. Lett.} {\bf 76} 877 (1996);
S.~Baek, P.~Ko, J.~Lee and H.S.~Song, 
{\it Phys. Lett.} {\bf B 389} 609 (1996).

\bibitem{lep6} P.~Cho, {\it Phys. Lett.} {\bf B 368} 171 (1996).

\bibitem{trans} P.~Cho and M.~Wise, {\it Phys. Lett.}
{\bf B 346} 129 (1995);
S.~Baek, P.~Ko, J.~Lee and H.S.~Song, 
{\it Phys.Rev.} {\bf D 55} 6839 (1997); 
M.~Beneke and I.~Rothstein, {\it Phys. Lett.} {\bf B 372} 157 (1996); 
M.~Beneke and M.~Kr\" amer, {\it Phys. Rev.} {\bf D 55} 5269 (1997).

\bibitem{hadro} S.~Gupta and K.~Sridhar, {\it Phys. Rev.} {\bf D 54} 5545 
(1996); 
M.~Beneke and I.~Rothstein, {\it Phys. Rev.} {\bf D 54} 2005 (1996); 
S.~Gupta and K.~Sridhar, {\it Phys. Rev.} {\bf D 55} 2650 (1997).

\bibitem{cakr} 
N.~Cacciari and M.~Kr\"amer, 
{\it Phys. Rev. Lett.} {\bf 76} 4128 (1996);
P.~Ko, J.~Lee, H.S.~Song, 
{\it Phys. Rev.} {\bf D 54} 4312 (1996).

\bibitem{brw} M. Beneke, I.Z. Rothstein, M.B. Wise,  
{\it Phys. Lett.} {\bf B 408} 373 (1997).

\bibitem{sms} K.~Sridhar, A.D.~Martin and W.J.~Stirling, 
{\it Phys. Lett.} {\bf B 438} 211 (1998).

\bibitem{lhc} K.~Sridhar, {\it Mod. Phys. Lett.} {\bf A 11} 1555 
(1996).

\bibitem{ours} K.~Sridhar, {\it Phys. Rev. Lett.} {\bf 77} 4880 (1996);
P.~Mathews, P.~Poulose and K.~Sridhar, {\it Phys. Lett.} {\bf B 438} 
336 (1998).

\bibitem{mangano} A.~Petrelli et al., 
{\it Nucl. Phys.} {\bf B514} 215 (1998); F.~Maltoni, M.L.~Mangano and
A.~Petrelli, {\it Nucl Phys.} {\bf B519} 361 (1998).
 
\bibitem{psigam} M.~Drees and C.S.~Kim, {\it Z.~Phys.} {\bf C 53} 673 (1992);
K.~Sridhar, {\it Phys. Rev. Lett.} {\bf 70} 1747 (1993); C.S.~Kim and 
E.~Reya, {\it Phys. Lett.}  {\bf B 300} 298 (1993).

\bibitem{kim} C.S.~Kim, J.~Lee, H.S.~Song, 
{\it Phys. Rev.} {\bf D 55} 5429 (1997).

\bibitem{us} D.P.Roy and K.~Sridhar, {\it Phys. Lett.} {\bf B 341} 413 
(1995).

\bibitem{mrs} A.D.~Martin, R.G.~Roberts and W.J.~Stirling,
{\it Phys. Lett.} B {\bf 306} 145 (1993);
{\it Phys. Lett.} B {\bf 309} 492 (1993).

\bibitem{ccsl} 
   B. Cano-Coloma and M.A. Sanchis-Lozano, {\it Nucl. Phys. } {\bf B508} 
753 (1997).

\bibitem{kniehl} B.A.~Kniehl and G.~Kramer, {\it Eur. Phys. J.} {\bf C6}
493 (1998).

\end{thebibliography}
\end{document}